\documentclass[%
preprint,
 amsmath,amssymb,
 aps,
]{revtex4-1}

\usepackage{dcolumn}
\usepackage{amsmath}
\usepackage{amssymb}
\usepackage{slashed}
\usepackage{hyperref}
\usepackage{graphicx}
\usepackage{feynmp-auto}
\usepackage{mathrsfs}

\usepackage{bm}

\def\p{\mbox{\boldmath$\displaystyle\mathbf{p}$}}

\def\0{\mbox{\boldmath$\displaystyle\mathbf{0}$}}

\def\x{\mbox{\boldmath$\displaystyle\mathbf{x}$}}
\def\y{\mbox{\boldmath$\displaystyle\mathbf{y}$}}

\newcommand{\dual}[1]{\overset{{}^{{}^{\boldsymbol{\neg}}}}{\smash[t]{#1}}} 

\begin{document}


\title{Constraints on mass dimension one fermionic\\
 dark matter from the Yukawa interaction}

\author{Marco Dias}
\email{mfedias@gmail.com}
\affiliation{Departamento de Ci\^{e}ncias Exatas e da Terra,\\
Universidade Federal de S\~{a}o Paulo Diadema, 09972-270 S\~{a}o Paulo, Brasil}

\author{Cheng-Yang Lee}
\email{cylee@ime.unicamp.br}
\affiliation{
Institute of Mathematics, Statistics and Scientific Computation,\\
Unicamp, 13083-859 Campinas, S\~{a}o Paulo, Brazil
}%

\date{\today}

\begin{abstract}
We study the loop corrections to the scalar propagator and the fermionic self-energy for the mass dimension one fermionic dark matter with the Yukawa interaction. We find, in the former case, there is a non-vanishing Lorentz-violating term while the later is Lorentz-invariant.  Our study of the fermionic loop correction shows that unitarity demands the fermionic mass must be at least half of the bosonic mass and that the Lorentz-violating term makes a non-trivial correction to the bosonic propagator. We discuss what these results mean in the context of the Standard Model and the possibility of bypassing the unitarity constraint. In the simplest scenario, within the framework of standard quantum field theory, by identifying the scalar boson to be the Higgs boson with a mass of 125 GeV, the mass of the fermion must be at least 62.5 GeV. 
\begin{description}
\item[PACS numbers]11.30.Cp, 12.60.Fr
\end{description}
\end{abstract}

\maketitle

\section{Introduction}

The mass dimension one fermionic field has many intriguing features~\cite{Ahluwalia:2004ab,Ahluwalia:2004sz,Ahluwalia:2008xi,Ahluwalia:2009rh}. 
Among them, the important features that characterize the theory are that the field satisfies the Klein-Gordon but not the Dirac equation and is of mass dimension one instead of three-half. Therefore, whatever these particles are, they must be physically distinct from the Dirac fermion and provide a promising direction of research for physics beyond the Standard Model (SM). Since its conception, the theory has been studied in various disciplines ranging from black hole~\cite{daRocha:2014dla,Cavalcanti:2015nna}, cosmology~\cite{
Boehmer:2006qq,Boehmer:2007ut,Boehmer:2007dh,Boehmer:2008rz,Boehmer:2008ah,Gredat:2008qf,Boehmer:2009aw,
Shankaranarayanan:2009sz,Boehmer:2010ma,Boehmer:2010tv,Chee:2010ju,Wei:2010ad,Shankaranarayanan:2010st,
S.:2014dja,Pereira:2014wta,Basak:2014qea}, mathematical physics~\cite{HoffdaSilva:2009is,daRocha:2008we,daRocha:2007pz,daRocha:2005ti,daRocha:2011yr,daRocha:2011xb,
daRocha:2013qhu,Cavalcanti:2014wia,Cavalcanti:2014uta,Bonora:2014dfa,Ablamowicz:2014rpa} and quantum field theory~\cite{Fabbri:2009aj,Fabbri:2009ka,Dias:2010aa,Fabbri:2010va,Ahluwalia:2010zn,Lee:2012td,
Alves:2014kta,Alves:2014qua,Agarwal:2014oaa,Lee:2015jpa}. For a comprehensive review on the subject, please see~\cite{Ahluwalia:2013uxa}.

Initial investigation revealed that the mass dimension one fermions have an intrinsic darkness with respect to the SM thus making them natural dark matter candidates~\cite{Ahluwalia:2004ab,Ahluwalia:2004sz}. What made the construction possible, bypassing the uniqueness of the Dirac field is that the theory does not satisfy Lorentz symmetry. Instead, it satisfies the symmetry of boost and rotation along a preferred direction. One may argue that Lorentz violation invalidates the theory but this is not necessarily true. While there are stringent constraints on Lorentz-violating theories~\cite{Mattingly:2005re,Ackermann:2009aa}, they are all confined to the SM sector. Currently, there is no direct evidence suggesting that dark matter satisfies Lorentz symmetry. 
Additionally, the fermionic field has a positive-definite free Hamiltonian, local interactions and furnishes fermionic statistics~\cite{Lee:2012td}. These properties are highly non-trivial and require careful choices of expansion coefficients and a field adjoint.

In this paper, we study the effects of Lorentz violation by computing the fermionic loop correction to the scalar propagator and the fermionic self-energy associated with the Yukawa interaction. The effects of loop-induced Lorentz-violation to the scalar propagator is non-zero whereas the fermionic self-energy is Lorentz-invariant. At one-loop, we find that unitarity, namely the optical theorem is violated unless the fermionic mass is at least half of the bosonic mass thus forbidding the decay of the scalar boson into a fermion anti-fermion pair. The Lorentz-violating term also makes a non-trivial correction to the bosonic propagator. We discuss what these results mean in the context of the SM and the possibility of bypassing the unitarity constraint.


\section{Loop corrections}

The theory under consideration here is the theory of mass dimension one fermion and a real scalar boson with the Yukawa interaction whose Lagrangian is
\begin{equation}
\mathscr{L}=\partial^{\mu}\dual{\Lambda}\partial_{\mu}\Lambda-m^{2}_{\Lambda}\dual{\Lambda}\Lambda-
g_{\phi}\dual{\Lambda}\Lambda\phi.
\end{equation}
In principle, one could also introduce interactions of the form $g'_{\phi}\dual{\Lambda}\Lambda\phi^{2}$ and $g_{\Phi}\dual{\Lambda}\Lambda\Phi^{\dag}\Phi$ where $\Phi(x)$ is a complex scalar field and $g'_{\phi}$ and $g_{\Phi}$ are dimensionless couplings. The reason why they are not considered here is that in this paper, our focus is on the loop corrections to the scalar propagator. Additionally, the fermionic loop for these interactions take the same form. The only differences are that $g_{\phi}$ has the dimension of mass and  that for the Yukawa interaction, the loop correction modifies the scalar propagator whereas the four-point interactions modify the vertices.

\begin{figure}
\begin{center}
\includegraphics[scale=0.3]{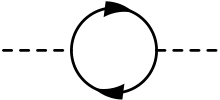}
\caption{Loop correction to the scalar propagator}\label{fig1}
\end{center}
\end{figure}

The fermionic loop of fig.~\ref{fig1} can be formally expressed as
$S_{\phi}(p)[-i(2\pi)^{4}\Pi^{*}_{\mbox{\tiny{1-loop}}}(p^{2})]S_{\phi}(p)$ where $S_{\phi}(p)$ is the free scalar propagator. We adopt the normalization where the free fermionic and scalar propagators are given by
\begin{eqnarray}
&& S_{\Lambda}(p)=\frac{i}{(2\pi)^{4}}\frac{I+\mathcal{G}_{p}}{p^{2}-m^{2}_{\Lambda}+i\epsilon},\\
&& S_{\phi}(p)=\frac{i}{(2\pi)^{4}}\frac{1}{p^{2}-m^{2}_{\phi}+i\epsilon}.
\end{eqnarray}
The matrix $\mathcal{G}_{p}$ is defined as
\begin{equation}
\mathcal{G}_{p}=i\left(\begin{matrix}
0 & 0 & 0 & -e^{-i\phi} \\
0 & 0 & e^{i\phi} & 0 \\
0 &-e^{-i\phi} & 0 & 0\\
e^{i\phi} & 0 & 0 & 0
\end{matrix}\right)
\end{equation}
with $\phi$ being the azimuthal angle defined by the momentum in the spherical coordinate
\begin{equation}
\p=|\p| (\cos\phi\sin\theta,\sin\phi\sin\theta,\cos\theta).
\end{equation}
The non-covariant fermionic propagator which contains information about the preferred direction is obtained by computing the time-ordered product $\langle\,\,|T[\Lambda(x)\dual{\Lambda}(y)]|\,\,\rangle$. The definition for $\Lambda(x)$ and $\dual{\Lambda}(y)$ are given in app.~\ref{A} and a detailed derivation of the propagator can be found in~\cite{Ahluwalia:2009rh}. 

Comparing the mass dimension one fermion to complex scalar bosons, apart from the spin-statistics, another important difference is that the fermionic propagator contains a non-covariant $\mathcal{G}_{p}$ matrix which is absent for its scalar counterpart. Unless one modifies the field adjoint $\dual{\Lambda}(x)$ as in ref.~\cite{Ahluwalia:2016rwl}, it is not possible to obtain a Klein-Gordon propagator. Therefore, the fermionic fields cannot be replaced by complex scalar fields.

Evaluating fig.~\ref{fig1} using the propagators, $\Pi^{*}_{\mbox{\tiny{1-loop}}}(p^{2})$ is given by
\begin{equation}
\Pi^{*}_{\mbox{\tiny{1-loop}}}(p^{2})=-2g^{2}_{\phi}\left[\frac{i}{(2\pi)^{4}}\right]
\int d^{4}k\frac{1}{(k^{2}-m^{2}_{\Lambda}+i\epsilon)[(k+p)^{2}-m^{2}_{\Lambda}+i\epsilon]}
+F(p)\label{eq:loop2}
\end{equation}
where $F(p)$ is defined as
\begin{equation}
F(p)\equiv -2g^{2}_{\phi}\left[\frac{i}{(2\pi)^{4}}\right]
\int d^{4}k\frac{\cos(\phi_{k}-\phi_{k+p})}{(k^{2}-m^{2}_{\Lambda}+i\epsilon)[(k+p)^{2}-m^{2}_{\Lambda}+i\epsilon]}.\label{eq:f}
\end{equation}
Equation~(\ref{eq:loop2}) is a sum of a Lorentz-invariant and Lorentz-violating integral. The former can be evaluated by the standard formalism of renormalization
\begin{equation}
\Pi^{*}_{\mbox{\tiny{1-loop}}}(p^{2})=-2g^{2}_{\phi}\left[\frac{i}{(2\pi)^{4}}\right]\int^{1}_{0}dx
(i\pi^{d/2})[m^{2}_{\Lambda}-p^{2}x(1-x)]^{d/2-2}\Gamma\left(2-\frac{2}{d}\right)+F(p).
\end{equation}
In the limit $d\rightarrow4$, we obtain
\begin{equation}
\Pi^{*}_{\mbox{\tiny{1-loop}}}(p^{2})=-\frac{g^{2}_{\phi}}{8\pi^{2}}\int^{1}_{0}dx
\left(\ln[m^{2}_{\Lambda}-p^{2}(1-x)x]+\frac{2}{d-4}+\gamma\right)+F(p). \label{eq:Pi}
\end{equation}
where $\gamma$ is the Euler constant. 
The trigonometric function in eq.~(\ref{eq:f}) can be expanded in terms of momenta to give us
\begin{eqnarray}
\cos(\phi_{k}-\phi_{k+p})&=&\cos\phi_{k}\cos\phi_{k+p}+\sin\phi_{k}\sin\phi_{k+p}\nonumber\\
&=&\frac{k_{x}(k_{x}+p_{x})+k_{y}(k_{y}+p_{y})}{\sqrt{(k_{x}^{2}+k_{y}^{2})[(k_{x}+p_{x})^{2}+(k_{y}+p_{y})^{2}]}}.\label{eq:cos}
\end{eqnarray}

From eq.~(\ref{eq:cos}), we see that $\Pi^{*}_{\mbox{\tiny{1-loop}}}(p^{2})$ is a function of $p_{x}$ and $p_{y}$ so it is not Lorentz-invariant. But when $p_{x}=p_{y}=0$, the function $F(p)$ becomes identical to the first term so it can contribute as much as 50\% to $\Pi^{*}_{\mbox{\tiny{1-loop}}}(p^{2})$. However, aligning the momentum along the $z$-axis does not necessarily define a preferred frame or direction. For example, in the $\phi_{1}\phi_{2}\rightarrow\phi_{3}\phi_{4}$ scattering of the $\phi^{3}$ theory, there are non-zero contributions from the $s$, $t$ and $u$-channels. At one-loop (fig.~\ref{fig1}), the momenta associated with these channels are respectively given by $\p_{1}+\p_{2}$, $\p_{1}-\p_{3}$ and $\p_{1}-\p_{4}$ and it is impossible to choose a frame where all three momenta are vanishing. It follows that there exists no preferred frame or direction such that $\cos(\phi_{p}-\phi_{p+k})=1$ for all three channels.  

\subsection{The optical theorem and correction to the bosonic propagator}
\begin{figure}
\includegraphics[scale=0.4]{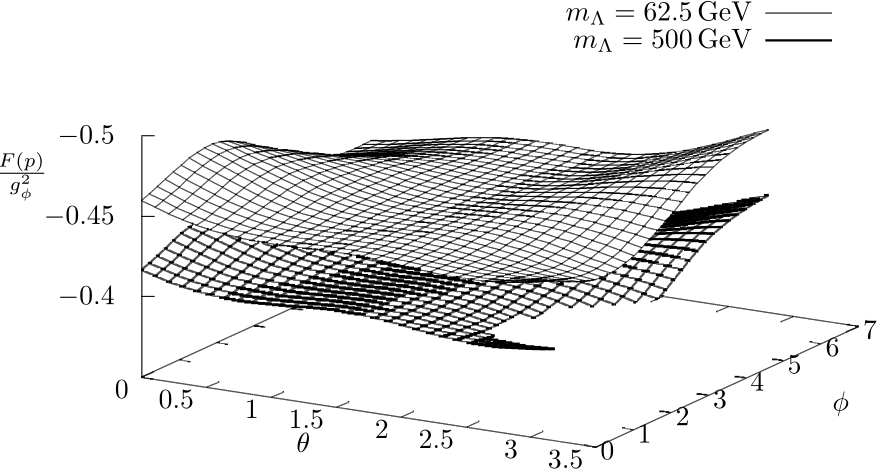}
 \label{Fp}
\caption{The plot of $F(p)/g^2_\phi$ obtained by Monte Carlo integration with $5.8\times 10^{7}$ sample points that represents the angular distribution in the $\theta-\phi$ plane. We considered $m_\Lambda=62.5\,\mbox{GeV},m_\Lambda=500\, \mbox{GeV}$ with an effective cut-off
$\mu_{\textnormal{eff}}=1.2209\times 10^{19}\mbox{GeV}$ at the Planck scale.}
\end{figure}
In standard quantum field theory, two of the most important concepts are Lorentz symmetry and unitarity. Since the theory under consideration only satisfies the symmetry of boost and rotation along a preferred direction instead of the full Lorentz group, it is instructive to check the unitarity of the theory. Towards this end, we make use of the optical theorem:
\begin{equation}
\mbox{Im}[\Pi_{\alpha\alpha}(m^{2}_{\alpha})]=-\frac{1}{2}\sum_{X}\Gamma(\alpha\rightarrow X)
\end{equation}
where $\alpha$ is a one-particle state, $\Pi_{\alpha\alpha}(m^{2}_{\alpha})$ is the two-point function and $\Gamma(\alpha\rightarrow X)$ being the total decay rate to some final multi-particle state $X$. In our case, $\Pi_{\alpha\alpha}(m^{2}_{\alpha})$ is the fermionic-loop correction to the scalar propagator given by eq.~(\ref{eq:loop2}) evaluated at $p^{2}=m^{2}_{\phi}$ and the decay rate is $\phi\rightarrow\dual{\Lambda}\Lambda$ is evaluated at tree-level. The later evaluates to (see app.~\ref{A})
\begin{equation}
\Gamma(\phi\rightarrow\dual{\Lambda}\Lambda)=\frac{g^{2}_{\phi}}{2\pi m_{\phi}}
\sqrt{1-\frac{4m^{2}_{\Lambda}}{m^{2}_{\phi}}}\left(\frac{m^{2}_{\phi}}{2m^{2}_{\Lambda}}-1\right).\label{eq:decay}
\end{equation}
For the decay to occur, one requires $m_{\phi}>2m_{\Lambda}$. When this condition is satisfied, the Lorentz-invariant integral of $\Pi^{*}_{\mbox{\tiny{1-loop}}}(m^{2}_{\phi})$ has an imaginary part. 
Here, to simplify the problem, we can avoid in having to evaluate $F(p)$ by taking the external momentum to be $p_{x}=p_{y}=0$. The argument of the natural logarithm in eq.~(\ref{eq:Pi}) is negative in the range $x\in[x_{-},x_{+}]$ where
\begin{equation}
x_{\pm}=\frac{1}{2}\left[1\pm\sqrt{1-\frac{4m^{2}_{\Lambda}}{m^{2}_{\phi}}}\right]
\end{equation}
so the imaginary part of $\Pi^{*}_{\mbox{\tiny{1-loop}}}(m^{2}_{\phi})$ is given by
\begin{equation}
\mbox{Im}[\Pi^{*}_{\mbox{\tiny{1-loop}}}(m^{2}_{\phi})]=
-\frac{g^{2}_{\phi}}{4\pi m_{\phi}}\sqrt{1-\frac{4m^{2}_{\Lambda}}{m^{2}_{\phi}}}.\label{eq:imaginary}
\end{equation}

Comparing eqs.~(\ref{eq:decay}) and (\ref{eq:imaginary}), we find that for $m_{\phi}>2m_{\Lambda}$ the optical theorem and hence unitarity is violated. If we wish to preserve unitarity, we must have $m_{\Lambda}\geq \frac{1}{2}m_{\phi}$ so that the decay channel $\phi\rightarrow\dual{\Lambda}\Lambda$ is forbidden. However, this seems unnatural since we should expect the optical theorem to hold for all ranges of masses. But in our opinion, within the standard framework of quantum field theory, this is an inevitable consequence the theory has to confront. Lorentz violation does not provide an exception to the optical theorem as its derivation only assume unitarity of the $S$-matrix and not the underlying space-time symmetry. A proposal to bypass this problem has recently been proposed in~\cite{Ahluwalia:2016rwl}, the details will be discussed in the conclusion.

In this paper, we work within the standard framework of quantum field theory so we must impose the inequality $m_{\phi}>2m_{\Lambda}$. Apart from this, another important issue that deserves our attention is the corrections to the scalar propagator which we now consider. For this purpose, it is instructive to introduce an effective cut-off $\mu_{\mbox{\tiny{eff}}}$. After performing the Feynman parametrization, Wick rotation, we get
\begin{equation}
\Pi^{*}_{\tiny{\mbox{1-loop}}}(p^{2})=\frac{g^{2}_{\phi}}{8\pi^{2}}\int^{1}_{0}dx
\ln\left[\frac{\mu_{\mbox{\tiny{eff}}}^{2}}{m^{2}_{\Lambda}-m^{2}_{\phi}(1-x)x}\right]
+F(p)+O(g^{4}_{\phi})
\label{eq:delta_m}
\end{equation}
where the same cut-off is also applied to $F(p)$. 
To evaluate $F(p)$, the Feynman parametrization and Wick rotation become inconvenient. Instead, we perform the $k^{0}$ integral analytically, and then evaluate the rest using Monte Carlo integration. The integral for $F(p)$ can be expressed as
\begin{equation}
F(p)=\frac{g^{2}_{\phi}}{8\pi^{3}}\int d^{3}k \cos(\phi_{k}-\phi_{k+p})
\left\{\frac{1}{E_{k}[(E_{k}-E_{p})^{2}-E^{2}_{p+k}]}+\frac{1}{E_{p+k}[(E_{p+k}+E_{p})^{2}-E^{2}_{k}]}
\right\}
\end{equation}
where
\begin{eqnarray}
&& E_{k}=\sqrt{|\textbf{k}|^{2}+m^{2}_{\Lambda}},\\
&& E_{p}=\sqrt{|\p|^{2}+m^{2}_{\Lambda}},\\
&& E_{p+k}=\sqrt{|\p+\textbf{k}|^{2}+m^{2}_{\Lambda}}.
\end{eqnarray}
The integration is performed with $\p=(p_{x},p_{y},p_{z})$ and $m_{\Lambda}$ being the free parameters. Figure~\ref{Fp} provides a graphical representation of $F(p)/g^{2}_{\phi}$ with $m_{\Lambda}=62.5$ GeV, $m_{\Lambda}=500$ GeV and an effective cut-off at the Planck scale.  



There are two important consequences that need to be noted. Firstly, the magnitude of $F(p)$ within the considered domain is finite and smooth, showing no divergent behaviour even when the effective cut-off is taken to be the Planck scale. This suggests that both $F(p)$ and $F(p,m_{\phi})$ where the later is defined to be on-shell, are finite.  Therefore, we can be confident that at one-loop, the function $\Pi^{*}(p^{2})$, which is a sum of $\Pi^{*}_{\mbox{\tiny{1-loop}}}(p^{2})$ and the counter-terms is also finite. Specifically, the renormalization condition $\Pi^{*}(m^{2}_{\phi})=0$ cancels the momentum-independent divergent terms so that $\Pi^{*}(p^{2})$ is a sum of Lorentz-invariant functions and $F(p)-F(m^{2}_{\phi},p)$~\footnote{In this paper, we have adopted the on-shell subtraction scheme where the renormalized mass and the pole mass of the propagator are equal. The renormalization conditions are $\Pi^{*}(m^{2}_{\phi})=0$ and $d/dp^{2}\Pi^{*}(m^{2}_{\phi})=0$. Similarly, for the mass dimension one fermion, we have $\Sigma^{*}(m^{2}_{\Lambda})=0$ and $d/dp^{2}\Sigma^{*}(m^{2}_{\phi})=0$}. Secondly, from the form of the integral of $F(p)$, it is clear that this function does not provide a dominant contribution to $\Pi_{\tiny{\mbox{1-loop}}}(p)$. Nevertheless, the plot shows that the contribution has certain angular dependence. In the context of the SM, where the scalar boson is identified to be the Higgs boson, this means that the Higgs propagator is not Lorentz-invariant but it instead has non-zero fluctuations when measured in different directions. Therefore, any such fluctuation detected in physical processes involving the Higgs propagator could be an indirect evidence of the mass dimension one fermions.

The finiteness of $F(p)$ with an effective cut-off taken at the Planck scale suggest that although the theory violates Lorentz symmetry, it is power-counting renormalizable. In particular, the matrix elements of $\mathcal{G}_{p}$ which is responsible for the violation, do not increase with momentum. The theory does however, exhibits a non-local behaviour. A direct computation shows that the equal-time field-conjugate momentum anti-commutator is~\cite{Ahluwalia:2008xi,Ahluwalia:2009rh}
\begin{equation}
\{\Lambda(t,\x),\Pi(t,\y)\}=i\int\frac{d^{3}p}{(2\pi)^{3}}
e^{-i\mathbf{p\cdot(x-y)}}[I+\mathcal{G}_{p}]
\end{equation}
where it reduces to $i\delta^{3}(\x-\y)$ only when $\x-\y$ is aligned to the $z$-axis. The non-locality of the anti-commutator may be undesirable, but it captures the peculiar features of the theory. Although it prevents us from formulating the theory in the path-integral formalism, it does not stop us from constructing local interactions in the operator formalism since a direct evaluation shows that $\{\Lambda(t,\x),\dual{\Lambda}(t,\y)\}=O$. Therefore, as long as the interactions are functions of $g\dual{\Lambda}\mathcal{O}\Lambda$ where $\mathcal{O}$ is some local operator, causality will be preserved.

\begin{figure}
\begin{center}
\includegraphics[scale=0.3]{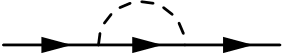}
\caption{Fermionic self-energy}\label{fig3}
\end{center}
\end{figure}

\subsection{Fermionic self-energy}

Now we consider fig.~\ref{fig3} whose expression in terms of the free fermionic propagator can be formally written as $S_{\Lambda}(p)[-i(2\pi)^{4}\Sigma^{*}_{\mbox{\tiny{1-loop}}}(p^{2})]S_{\Lambda}(p)$.
Evaluating the diagram, we obtain
\begin{equation}
\Sigma^{*}_{\mbox{\tiny{1-loop}}}(p^{2})=\frac{ig^{2}_{\phi}}{(2\pi)^{4}}\int d^{4}k
\left(\frac{1}{k^{2}-m^{2}_{\phi}+i\epsilon}\right)
\left[\frac{I+\mathcal{G}_{p-k}}{(p-k)^{2}-m^{2}_{\Lambda}+i\epsilon}\right].
\end{equation}
Shift the variable by $k\rightarrow k+p$ and take $d^{4}k=dk^{0} d^{3}k$ with $d^{3}k$ defined in the spherical coordinate, the integration over $\mathcal{G}_{k}$ identically vanishes leaving us with a Lorentz-invariant integral
\begin{equation}
\Sigma^{*}_{\mbox{\tiny{1-loop}}}(p^{2})=\frac{ig^{2}_{\phi}}{(2\pi)^{4}}\int d^{4}k
\left[\frac{1}{(k+p)^{2}-m^{2}_{\phi}+i\epsilon}\right]\left(\frac{I}{k^{2}-m^{2}_{\Lambda}+i\epsilon}\right).
\end{equation}
In the on-shell subtraction scheme where $m^{2}_{\Lambda}\equiv m^{2}_{\tiny{\mbox{bare}},\Lambda}-\delta m^{2}_{\Lambda}$ and 
$\Lambda_{\tiny{\mbox{bare}}}(x)\equiv Z^{1/2}\Lambda(x)$, the complete self-energy function $\Sigma^{*}(p^{2})$ is
\begin{equation}
\Sigma^{*}(p^{2})=-(Z-1)(p^{2}-m^{2}_{\Lambda})+Z\delta m^{2}_{\Lambda}+\Sigma^{*}_{\tiny{\mbox{1-loop}}}(p^{2}).
\end{equation}
Applying the renormalization condition $\Sigma^{*}(m^{2}_{\Lambda})=0$ and introducing an effective cut-off,
to the order $Z=1+O(g^{4}_{\phi})$, we get
\begin{equation}
\delta m^{2}_{\Lambda}=\frac{g^{2}_{\phi}}{16\pi^{2}}\int^{1}_{0}dx
\ln\left[\frac{\mu_{\mbox{\tiny{eff}}}^{2}}{m^{2}_{\Lambda}-(2m^{2}_{\Lambda}-m^{2}_{\phi})x+m^{2}_{\Lambda}x^{2}}
\right]+O(g^{4}_{\phi}).\label{eq:delta_ml}
\end{equation}
For $\delta m^{2}_{\Lambda}$ to be a real number, the denominator of the natural logarithm must be positive so we require the following inequality to hold
\begin{equation}
m^{2}_{\Lambda}-(2m^{2}_{\Lambda}-m^{2}_{\phi})x+m^{2}_{\Lambda}x^{2}\geq0
\end{equation}
for $x\in[0,1]$. Let $m^{2}_{\Lambda}=\alpha m^{2}_{\phi}$, the inequality becomes
$f(x)\geq0$
where 
\begin{equation}
f(x)=\alpha-(2\alpha-1)x+\alpha x^{2}.
\end{equation}
Since $\alpha>0$, the function has a minimum at $x_{0}=\frac{1}{2\alpha}(2\alpha-1)$. Therefore, the inequality is satisfied if $f(x_{0})\geq0$. This gives us the condition $\alpha\geq\frac{1}{4}$ which is in agreement with our earlier result.

\section{Conclusions}\label{conc}

In this paper, we have studied the simplest loop corrections for the mass dimension one fermionic dark matter with the Yukawa interaction. The Lorentz violation generated by the fermionic loop correction to the scalar propagator is non-zero and the fermionic self-energy is Lorentz-invariant. At one-loop, we find that the optical theorem is violated unless $m_{\Lambda}\geq\frac{1}{2}m_{\phi}$. 

Identifying the scalar boson to be the Higgs boson, the mass dimension one fermion must then be at least 62.5 GeV. The constraint on the bosonic and fermionic masses seems unnatural. Nevertheless, working within the standard framework of quantum field theory, this is an inevitable consequence that the theory has to confront. Unitarity violation may be a reason why the theory is inconsistent. But in our opinion, we should keep an open mind and exhaust all possibilities. At the same time, it is important to remind ourselves that there are no reasons why the masses cannot satisfy the required inequality.

One the most important results we have found is that if such fermions existed and interact with the Higgs boson via the Yukawa interaction, the Higgs propagator cannot be Lorentz-invariant. Instead, it is dependent on the frame of reference in which it is being computed. Therefore, any variations to the physical processes involving the Higgs propagator could indicate indirect evidence for the existence of mass dimension one fermions. 

Furthermore, the asymmetry in the Higgs mass in relation to the angles $\eta-\phi$ is a striking signature given by the current model for search in the nowadays accelerators. This is closely linked to the precision of the detectors to measure  the transverse energy $ E_t $ in the  angular plan. Future studies in this line can constrain the coupling constant in this model.



Finally, we should mention that a possible resolution to the unitarity problem has been proposed~\cite{Ahluwalia:2016rwl}. The resolution is based on the observation that the dual space of the spinors and the field adjoint of the theory are different from their Dirac counterpart. Their introduction ensured the locality of the fermionic fields and positivity of the free Hamiltonian. Given that they play such an important role, it was proposed that instead of using the Hermitian conjugation to compute the transition probability, one should use a new conjugation for processes involving mass dimension one fermions. Upon adopting the new conjugation, the fermionic propagator becomes the Klein-Gordon propagator. As a result, the fermionic loop correction is Lorentz-invariant and the optical theorem is satisfied. While this result is desirable, it should be noted that it is a departure from the standard quantum field theory and further investigation is needed to determine its mathematical consistency.

\section*{Acknowledgements}
MD is grateful for the resources provided by the Center for Scientific Computing (NCC/GridUNESP) of the S\~{a}o Paulo State University (UNESP). CYL is grateful to D.~V.~Ahluwalia and Giorgio Torrieri for useful discussions and  would like to thank the generous hospitality offered by IUCAA where part of this work was completed. CYL is supported by the CNPq grant 313285/2013-6.

\appendix

\section{Useful identities}\label{A}

The mass dimension one fermionic field operator and its adjoint are given by
\begin{eqnarray}
&&\Lambda(x)=(2\pi)^{-3/2}\int\frac{d^{3}p}{\sqrt{2mE_{\mathbf{p}}}}\sum_{\alpha}
[e^{-ip\cdot x}\xi(\p,\alpha)+e^{ip\cdot x}\zeta(\p,\alpha)b^{\dag}(\p,\alpha)],\\
&&\dual{\Lambda}(x)=(2\pi)^{-3/2}\int\frac{d^{3}p}{\sqrt{2mE_{\mathbf{p}}}}\sum_{\alpha}
[e^{ip\cdot x}\dual{\xi}(\p,\alpha)a(\p,\alpha)+e^{-ip\cdot x}\dual{\zeta}(\p,\alpha)b(\p,\alpha)].
\end{eqnarray}
The spinors $\xi(\p,\alpha)$ and $\zeta(\p,\alpha)$ are eigenspinors of the charge conjugation operator 
\begin{equation}
\mathcal{C}\xi(\p,\alpha)=\xi(\p,\alpha),\hspace{0.5cm}
\mathcal{C}\zeta(\p,\alpha)=-\zeta(\p,\alpha)
\end{equation}
where 
\begin{equation}
\mathcal{C}=\left(
\begin{matrix}
O & -i\Theta \\
i\Theta & O
\end{matrix}\right)K,\hspace{0.5cm}
\Theta=\left(
\begin{matrix}
0 & -1 \\
1 & 0
\end{matrix}\right)
\end{equation}
with $K$ being the complex-conjugation operator. The dual spinors is defined as
\begin{equation}
\dual{\xi}(\p,\alpha)=\overline{\xi}(\p,\alpha)\Xi(\p),\hspace{0.5cm}
\dual{\zeta}(\p,\alpha)=\overline{\zeta}(\p,\alpha)\Xi(\p).
\end{equation}
where
\begin{equation}
\Xi(\p)=\frac{1}{m}\sum_{\alpha}[\xi(\p,\alpha)\overline{\xi}(\p,\alpha)-\zeta(\p,\alpha)
\overline{\zeta}(\p,\alpha)].
\end{equation}
The spin-sums needed to compute the $\phi\rightarrow\dual{\Lambda}\Lambda$ decay rate are
\begin{eqnarray}
&&\sum_{\alpha}\xi(\p,\alpha)\overline{\xi}(\p,\alpha)=\slashed{p}[I+\mathcal{G}(\phi)],\\
&&\sum_{\alpha}\zeta(\p,\alpha)\overline{\zeta}(\p,\alpha)=\slashed{p}[I-\mathcal{G}(\phi)].
\end{eqnarray}

\bibliography{Bibliography}
 \bibliographystyle{unsrt}


\end{document}